\documentclass{aa}
\usepackage{graphicx,epsfig,rotating,amsmath,epsf,./txfonts,natbib}
\def\kms{\rm km \;s$^{-1}$}
\def\arcsec{$^{\prime\prime}$}

\begin{document}

\title{Coronal hole boundaries evolution at small scales: I. EIT 195 \AA\  and TRACE 171 \AA\ view}
\author{ M.~S. Madjarska\inst{1, 2} \and T. Wiegelmann\inst{1}}
\offprints{M.S. Madjarska, madj@arm.ac.uk}
\institute{Max-Planck-Institut f\"ur Sonnensystemforschung, Max-Planck-Str.
2, 37191 Katlenburg-Lindau, Germany \and Armagh Observatory, College Hill, Armagh BT61 9DG, N. Ireland}
           
\date{Received date, accepted date}
\abstract
{}
{We aim at studying the small-scale evolution at the  boundaries of an equatorial coronal hole 
connected with a channel of open magnetic flux with the polar region and an `isolated' one 
in the extreme-ultraviolet spectral range. We intend to determine the spatial and temporal scale of 
these changes. }
{Imager data from TRACE in the Fe~{\sc ix/x}~171~\AA\  
passband  and EIT on-board Solar and 
Heliospheric Observatory in the Fe~{\sc xii}~195 \AA\ passband were analysed.}
{We found that small-scale loops known as bright points play an essential role in coronal holes 
boundaries evolution at small scales. Their emergence and disappearance continuously  
expand or contract  coronal holes. The changes appear to be random on a time scale comparable with the lifetime 
of the loops seen at these temperatures. No signature was found for a major 
energy release during the evolution of the loops.}
{Although coronal holes seem to maintain their general shape during a few solar rotations, 
a closer look at their day-by-day and even hour-by-hour evolution demonstrates 
 a significant dynamics. The small-scale loops (10\arcsec--40\arcsec and smaller) which are abundant 
 along coronal hole boundaries  have  a  contribution to the small-scale evolution of coronal holes.
  Continuous magnetic reconnection of the open magnetic field lines of the coronal hole and the 
  closed field lines of the loops in the quiet Sun is more likely to take place.}


\authorrunning{Madjarska \& Wiegelmann}
\titlerunning{Coronal hole boundaries evolution}

\maketitle
\section{Introduction}
\label{sec:intro}

Coronal holes (CHs) are large regions on the Sun that are magnetically open.
They are identified as the source of the fast solar wind ($\sim$800 \kms) \citep{krieger73} and
are visible in coronal  lines (formed at temperatures
above 6~10$^5$~K) as regions  with a reduced emission relative to the quiet Sun 
\citep{wilhelm2000, stucki2002}. There are two types of coronal
holes: polar and mid-latitude CHs. During the minimum of solar activity  the 
solar atmosphere is dominated by two large CHs situated at both polar regions. The 
mid-latitude CHs can  be  either `isolated' or  connected  with a  channel of 
open magnetic flux to a polar CH. The latter are called  equatorial extensions 
of polar CHs (EECHs).  The isolated coronal holes have an occurrence rate that follows 
the solar activity cycle and are usually connected with an active region \citep{insley95}. 

\begin{figure*}[ht!]
\centering
\vspace{10cm}
        \includegraphics{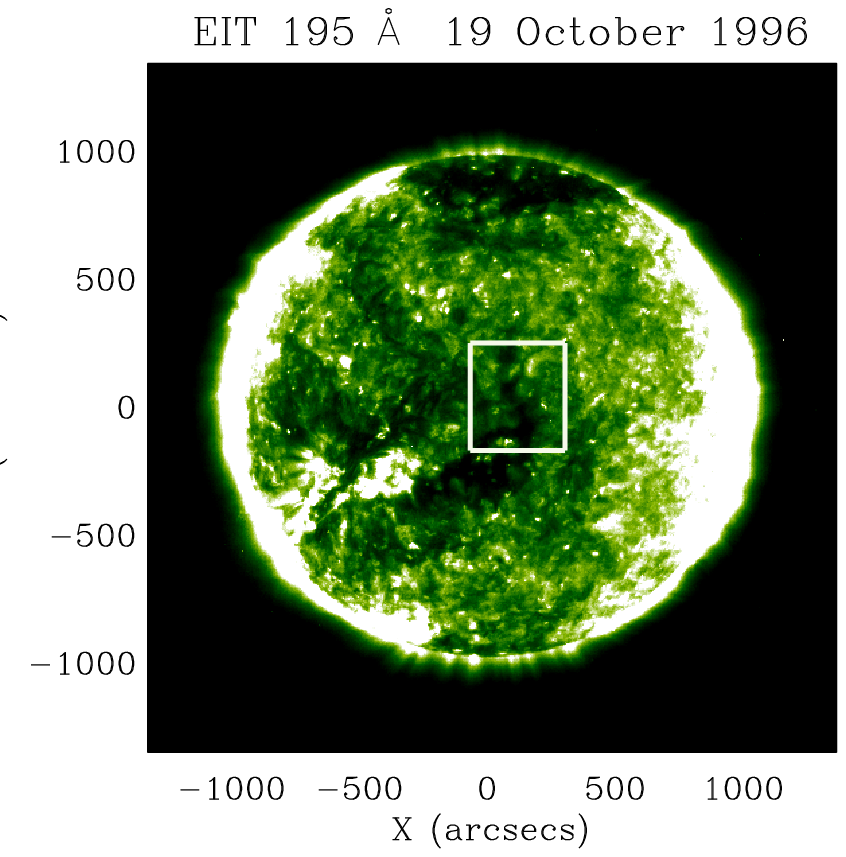}
	\includegraphics{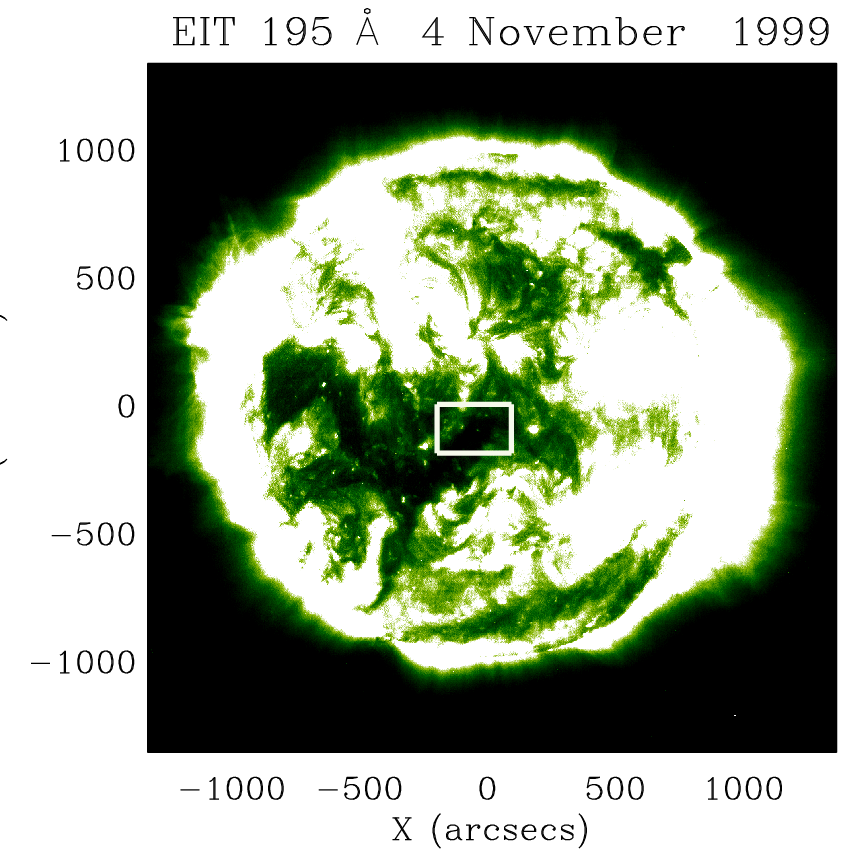}
	\caption{Full-disk EIT 195 \AA\ images obtained on 1996 October 19 and 1999 November 4.  
	The over-plotted boxes outline the field-of-views subject to a detailed study.}
	\label{fig1}
\end{figure*}

\cite{huber74}  compared the appearance and physical parameters of the different layers of the solar atmosphere in and outside 
coronal holes (e.g. the quiet Sun) using the Apollo Telescope Mount on Skylab.  Their measurements of the height of emission of 
various ions at different ionisation stages (different formation temperatures) at the polar limb
 indicated an increase of the thickness of the transition region underlying coronal holes.
Hence, they found a difference between quiet Sun and coronal holes already pronounced at transition region temperatures.

 \citet{feldman99} studied
 the morphology of the upper solar atmosphere using high-resolution data (1\arcsec -- 2\arcsec) taken 
 by Transition Region and Corona Explorer (TRACE) and Solar Ultraviolet Measurements of Emitted Radiation
   spectrometer (SUMER) on-board {\sc SoHO}, and the Naval Research Laboratory 
 spectrometer on {\sc Skylab}. The authors 
 found that in the temperature range 4~10$^4$~K ~$\le$ T$_e$ $\le$ ~1.4~10$^6$~K the upper solar
  atmosphere is filled with loops of different sizes with hotter 
and longer loops overlying the cooler and shorter loops \citep{dowdy86}. At heights above 2.5~10$^4$~km 
in the upper solar atmosphere of the quiet Sun only loops at temperatures T$_e$~$\sim$~1.4 
$\times$ 10$^6$ K exist. No distinction was found between quiet Sun and coronal hole morphology 
at 5~10$^4$ $\le$ T$_e$ $\le$  2.6~10$^5$ K. That suggests that both regions 
are filled with structures at similar sizes which are emitting at similar temperatures. 
These structures do not exceed a height of 7~Mm and have lengths $\le$~21 Mm. \citet{feldman99}
also investigated  the coronal hole boundaries concluding that they are seeded with 
small-scale loops ($<$~7~Mm). There coexist, however, long loops at temperatures above 
 T~$\sim$~1.4~10$^6$~K which generally originate from the same location but close at 
 faraway locations.   

\citet{thomas04} made a  further step by  reconstructing the  magnetic 
field in coronal holes and the quiet Sun with the help of a potential field model.
 They found that the CHs loops are on average shorter, lower and flatter than in the QS. 
High and long closed loops are extremely rare in  CHs, whereas short and low loops are almost
abundant as in the QS region. This result strongly  supports the observational findings 
on the structure of the  upper solar atmosphere in the quiet Sun and coronal holes.


\begin{figure*}[ht!]
\centering
\vspace{8cm}
        \includegraphics{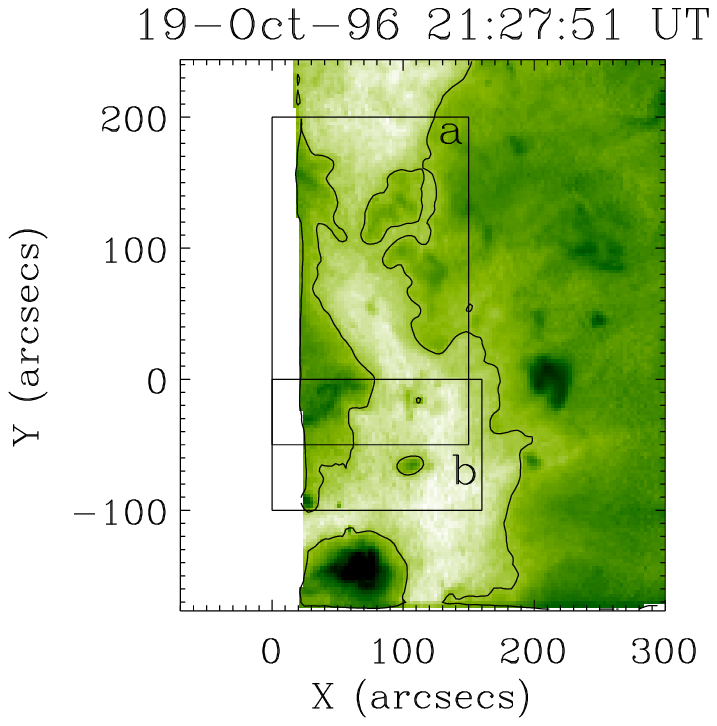}
	\includegraphics{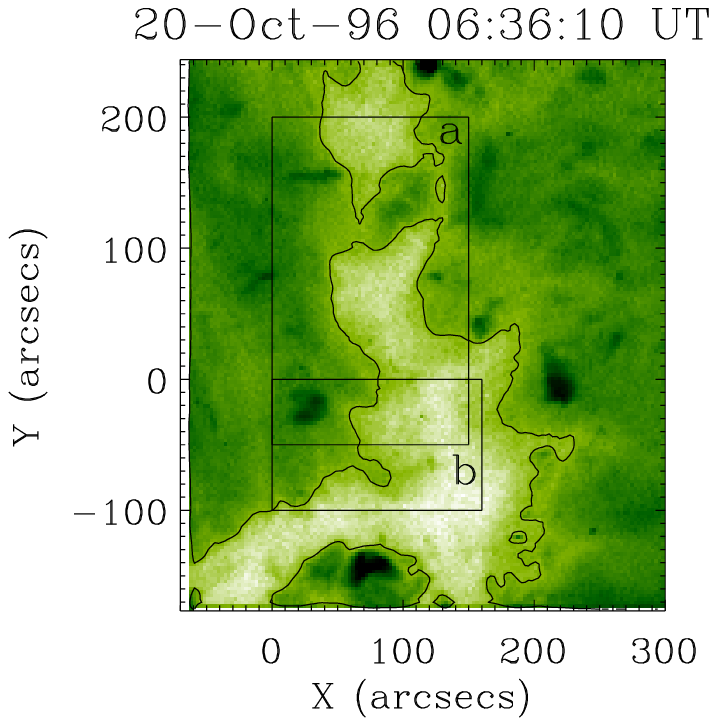}

	\caption{Color table  reversed EIT 195 \AA\ images  of the coronal hole observed on October 
	19-20, 1996. The over-plotted rectangular areas (a and b) are shown enlarged in 
	Figs.~3 and 4.}
	\label{fig2}
	\end{figure*}

Based on  {\sc Skylab} data in soft X-rays \citet{tim75} first reported the 
distinctive  feature of the EECHs to exhibit quasi-rigid rotation. \citet{shelke85} found that equatorial CHs 
(without dividing them in different classes) have a rotational 
period which is a function of latitude and thus exhibit differential rotation. 
\citet{navarro94} showed that isolated CHs have a typical 
differential rotation, while  EECHs maintain two types of rotation rates: 
differential below a  latitude of  40\degr\ which becomes almost rigid while approaching 
the poles. \citet{insley95} concluded that mid-latitude CHs rotate more 
rigidly than the photosphere, but still exhibit significant differential rotation.

Due to the different rotation profiles at coronal and photospheric level 
 and  the fact  that CH boundaries (CHBs) separate two topologically different 
(open and closed) magnetic field configurations, CHBs are  presumably the regions where 
  processes take place that open and close magnetic field  lines. The reconfiguration of the CHBs 
is believed to happen through  magnetic reconnection between the open and closed magnetic 
field lines of the CH and the surrounding  quiet Sun. 

The small-scale evolution of  CH boundaries has been a subject of several studies during the last few 
decades \citep[and the references therein]{kahler90}. These studies aimed at understanding 
the general evolution of coronal holes, the  quasi-rigid rotation of EECH compared  to the 
differentially rotating solar photosphere as well as their relation to the slow solar wind 
generation. \citet{kahler90} studied {\sc Skylab} X-ray  images of a single  EECH
with a time resolution of 90 minutes aimed at looking for discrete changes of the CH boundaries. 
They found that X-ray bright points (BPs, small-scale loops in the quiet Sun and CHs, 
for details see \citet{madj03}) play an important role in both the expansion and 
contraction of the CH. For this study a  single coronal hole observed during 
the decreasing phase of the solar cycle activity was used. \citet{bromage2000} reported that 
small-scale changes of the boundaries of the CH on 1996 August 26 took place  on time-scales of 
a few hours in observations with the  EIT in Fe~{\sc xii} 195~\AA. The authors, however, 
do not give any details on the nature of these changes. \citet {kahler02} made the 
first systematic morphological study of the boundaries of coronal holes as viewed in soft 
X-ray images from the Yohkoh Soft X-ray telescope. They studied three coronal holes during 
several rotations, all formed during the maximum of  solar activity cycles 
23 and 24. All three coronal holes represent equatorial extensions of 
polar coronal holes. They found that the CHs evolve slowly, and neither large-scale 
transient X-ray events nor coronal bright points appeared significant factors 
in the development of  CH boundaries. They suggested that open-close magnetic field reconnection 
is more likely to describe the actual physics at the CHBs. We should note here that the 
visibility of BPs  in X-ray images is 
strongly diminished during solar maximum activity by the presence of active regions and bright loops seen in X-rays. 
The number and size of BPs, and also  their lifetime strongly depend on the temperature 
of which they are observed, with more and longer living BPs seen in spectral lines 
formed at lower temperatures. 

 The emergence of magnetic fields in active regions and their subsequent diffusion due to random 
 convective motions in the photosphere influenced by the meridional flow is believed to be the main mechanism 
leading to the formation of coronal holes. \citet{wang94}  developed a model where the footpoints 
exchange between open and closed magnetic field lines (the so called `interchange reconnection') 
generates coronal holes and maintains their quasi-rigid rotation against the differentially 
rotating photospheric layers. This type of reconnection process results in an exchange of footpoints between open
and closed magnetic field lines with no change in the total amount of open or closed flux \citep{wang94}. 
The reconnection takes place very high in the corona ($r \sim  2.5$R$_{\sun}$) and occurs 
continuously in the form of small, stepwise displacements of field lines. According to the 
authors this scenario may explain why no X-ray signatures have 
been detected in association with boundary evolution in long-lived coronal holes studied by 
\citet{kahler02}. They proposed that the blobs emitted from the tops of  helmet
streamers as radially elongated density enhancements and associated with the slow solar wind 
are probably the result of this reconnection process.

Our study aims at providing for the first time an analysis on the small-scale 
(determined by the instrument resolution) evolution of coronal hole boundaries  on a  
time scale from tens of minutes to hours using EUV (EIT/195 \AA\ and TRACE/171 \AA) 
observations with spatial resolution ranging from 1\arcsec\ (TRACE) to 5.5\arcsec\ (EIT).
In Section~2 we describe the observational material and data reduction applied. 
Section~3 presents the data analysis and  the obtained results. The discussion and 
conclusions on the obtained results in the light of the existing theoretical models 
as well as the future perspectives on the subject discussed in this work are given in Section~4.

\section{Observational material and data reduction}

We studied two coronal holes, one was observed in October
 1996 (hereafter CH1) and the second in November 1999 (hereafter CH2). The CH1 is the well 
known `Elephant trunk coronal hole', an equatorial extension of north pole CH which has  been
the subject of several studies \citep[and the references therein]{bromage2000}. 
CH2 is an `isolated' mid-latitude CH observed in November  1999. Fig.~\ref{fig1}  shows 
EIT images in the 195~\AA\ passband  with over-plotted boxes outlining the analysed FOVs for both CHs.

\begin{figure*}[htp!]
\hspace{+1cm}
        \includegraphics{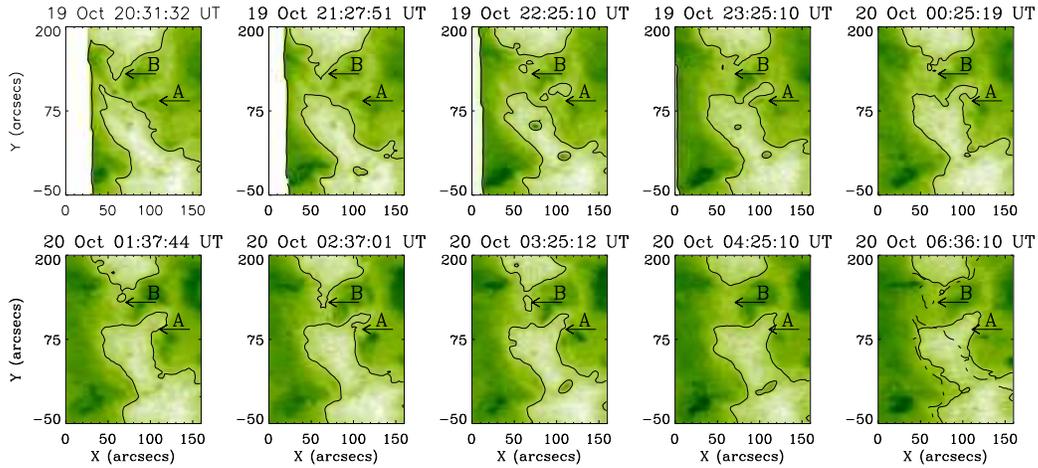}
	\caption{ EIT images of the area marked with  `a' in Fig.~\ref{fig2} (reversed color table).
	 The arrows point at the BPs whose evolution led to the expansion 
	 (arrow A) and the contraction (arrow B) of the CH. The over-plotted dashed-dotted line shows the boundary
	 from the first image. 
} 	\label{fig3}
\end{figure*}

\begin{figure*}[ht!]
\hspace{1cm}
        \includegraphics{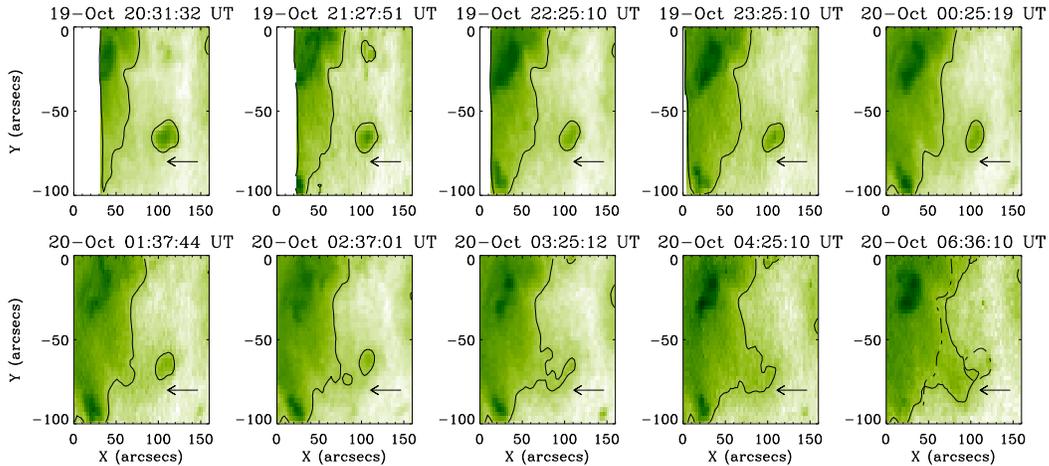}
	\caption{ EIT images of the area marked with  `b' in Fig~\ref{fig2} (reversed color table). 
	The arrows point at a bright point whose evolution led to the contraction of the 
	coronal hole. The over-plotted dashed-dotted line shows the boundary
	 from the first image. 
}
	\label{fig4}
\end{figure*}

\subsection{EIT}
The CH1 data were recorded with  EIT in the 195 \AA\ passband on 1996 October  19-20 with a 
full resolution of 2.6\arcsec. The images were taken 
from 20:00 UT (19 October) until 06:00 UT (20 October)  every 15~{\rm min}  (except at 05:00~UT). 
Additionally, we selected full-disk data covering 48 hrs of 
observations from 15:00~UT on October 18 until 15:41~UT  on October 20 taken  approximately every 2  hrs.
All data related to CH1 were derotated to 07:00~UT on October 20. The data from November
 1999 were taken from 15:00~UT  on November 3  
until 15:00~UT on November 5 and de-rotated to 15:00~UT on November 5. Both datasets have a 5.2\arcsec\ angular pixel size.

The necessary data reduction was applied, i.e. the images were background subtracted, degridded, 
flat-fielded, degradation corrected and normalised. They were also derotated to a 
reference time using the drot\_map.pro procedure from SolarSoft. The 
procedure applies synodic values for the differential rotation. Cosmic ray 
removal was not applied because it was not needed and carries the risk of removing real small-scale structures.

\begin{figure*}[thp!]
\centering\includegraphics{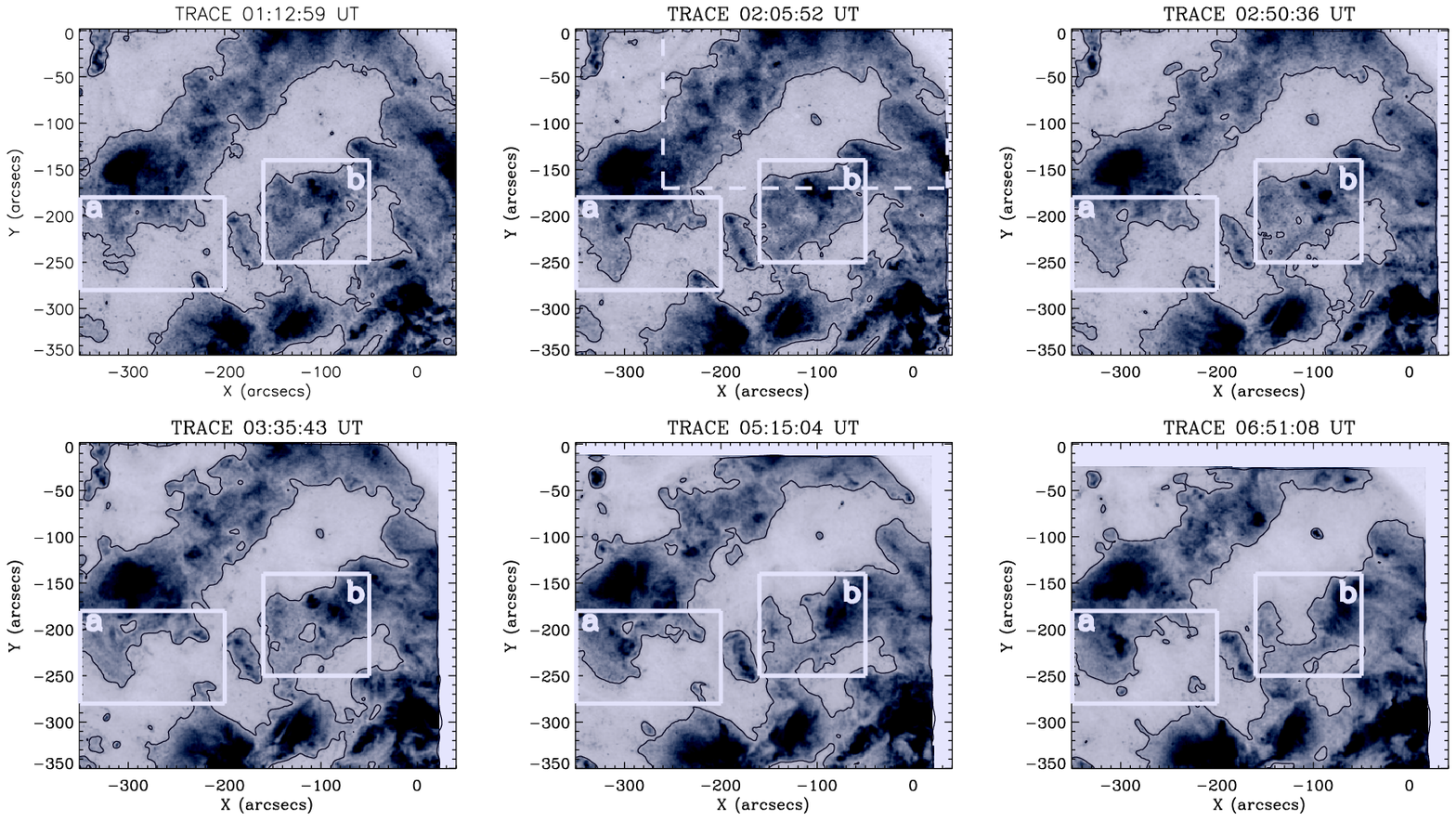}
	\caption{Reversed color TRACE 171 \AA\ images showing partial field-of-view of CH2 observed in
	November  1999.  The over-plotted  rectangular areas (a and b) are shown enlarged in  Figs 6 and 7. 
	The dashed line box corresponds to the field-of-view presented in Fig.~\ref{fig8}.}
	\label{fig5}
\end{figure*}

\subsection{TRACE}

The  TRACE data were obtained in  Fe~{\sc ix/x}~171 \AA\ on November 4, 1999. The data have a 3~{\sc s} 
cadence and a variable exposure time from 00:50~UT to 07:11~UT  during approximately 7 hours.
\begin{figure*}[htp!]
\centering
        \includegraphics[scale=0.8]{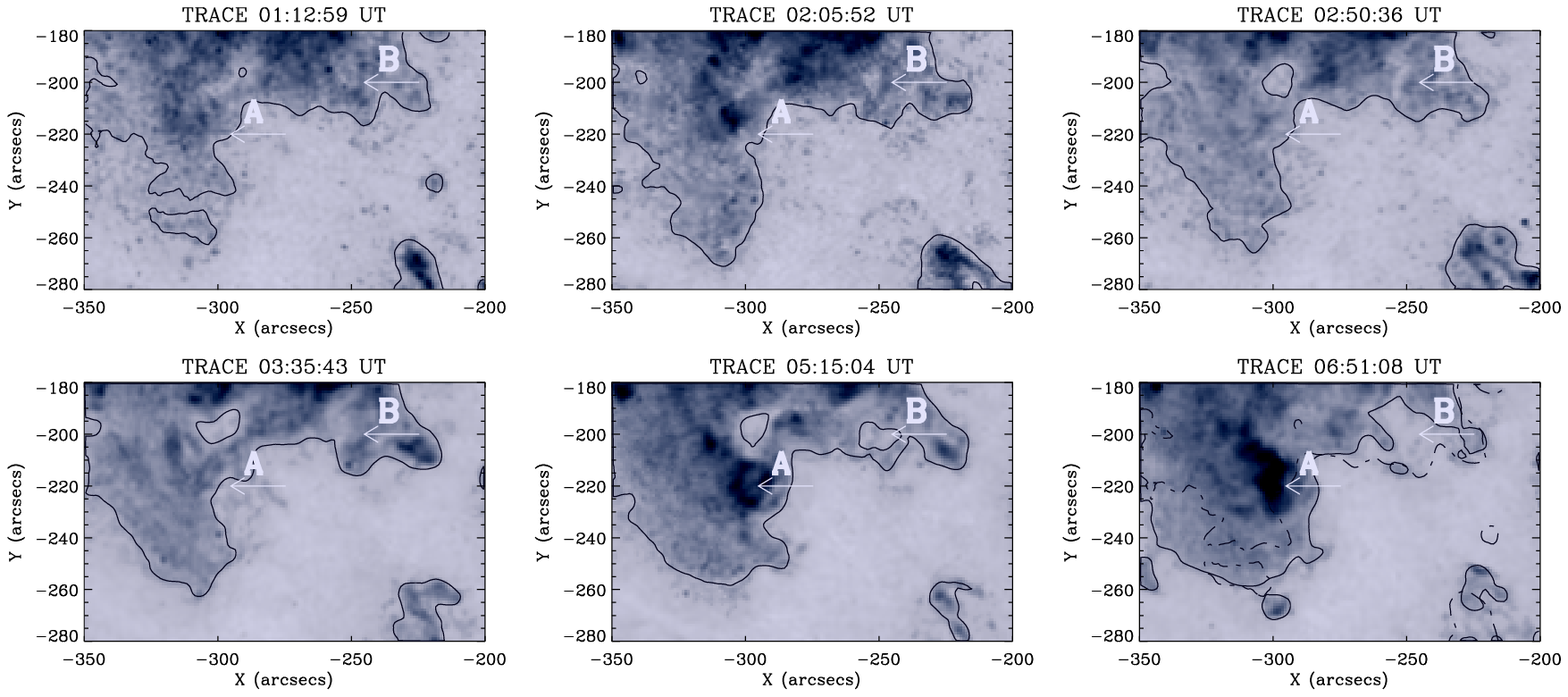}
	\caption{TRACE~171~\AA\  images marked with `a' in Fig. 5 and showing  a contraction of the CH 
	due to the formation of a new BP (arrow A) close to the CH boundary 
	 and an expansion due  to the disappearance of a BP (arrow B). The over-plotted dashed-dotted line shows the boundary
	 from the first image. }
	\label{fig6}
\end{figure*}


\begin{figure*}[htp!]
\centering

        \includegraphics[scale=0.8]{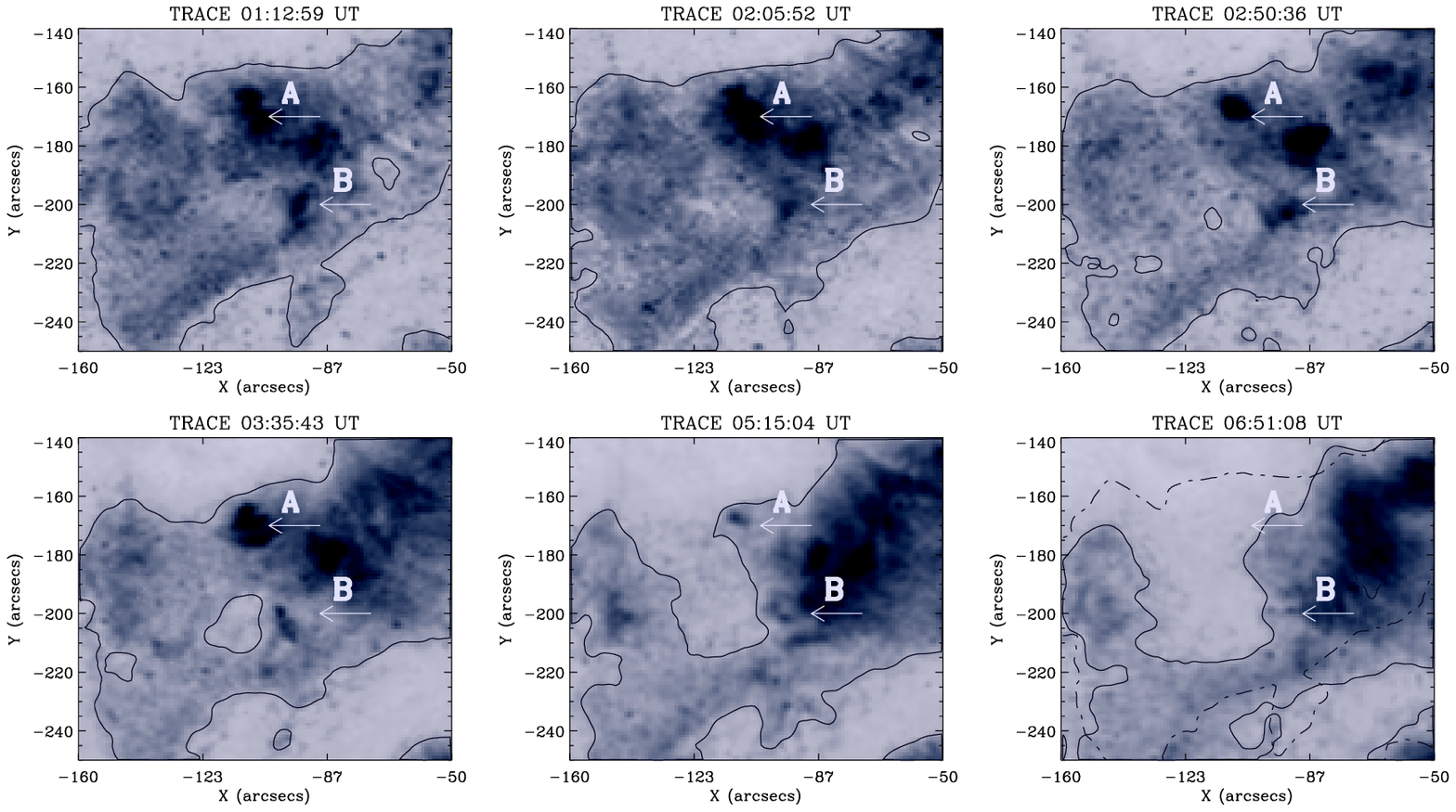}
	\caption{TRACE 171 \AA\ images marked with 'b' in Fig.~5  
	showing a large expansion of the coronal hole due to 
	the disappearance of two BPs (arrows A and B). The over-plotted dashed-dotted line shows the boundary
	 from the first image. }
	\label{fig7}
\end{figure*}

The following corrections were applied: dark current subtraction (obtained the day before),
flat-field corrections, spikes and streaks removal. The background diffraction pattern was 
removed but some residual remained. A normalisation for exposure and a pointing offset correction 
for the studied channel were also applied.

The data were obtained close to the eclipse period (which started on November 7)
 during which TRACE is attenuated by the Earth atmosphere in parts of its orbit, and 
 the response in the EUV channels is compromised. However, for several weeks  either side 
 of the eclipse period the data are also contaminated. As our data were taken only 3 days 
 before the eclipse started, we had to reduce  the number of used images  to only 6 (from 434). 
 The images were selected after a careful flux analysis and we are confident that  
 they can be used for the purposes of this study (see later in the text).

In both imagers (EIT and TRACE) the measured values are in data numbers (DN). Data number is the
 output of the instruments electronics  which corresponds to the incident photon signal 
  converted into charge within each CCD pixel. The contours were smoothed in order to remove changes due to 
  image irregularities which could be unreal changes.

\section{Data analysis and results}

In the present work we paid special attention on the data reduction in order to get reliable 
information on the evolution  of CH boundaries using a contour plot. This method is more 
reliable than a visual inspection and ensures the detection of any small-scale changes. The contours 
have different values for the data with different resolution. The contour  values also depend on
 the observed temperatures with coronal holes expanding at higher temperatures \citep{bromage2000}.   For the present 
 study we determined the boundaries as the region which has intensities 1.5 times the average intensity
  of the darkest region inside the coronal hole.   The visual inspection showed that this value describes well the boundaries
 as  determined using He~{\sc i} 10830~\AA\ images  from
 Kitt Peak National Observatory \citep{henney2005}.  For the TRACE~171~\AA\  images of CH2 we used an EIT image
   obtained in Fe~XII~195~\AA\ in order
  to establish the CHBs. The Fe IX/X~171~\AA\ passband has a transition region emission contribution 
  and is, therefore, not well suited to  determine CHBs. We selected two images (EIT and TRACE) obtained 
  at the same time, applied the necessary offset correction, over-plotted on  the TRACE images the boundaries as 
  determined by EIT and then established  for TRACE its own contour of constant flux which matches well the EIT contour.

In order to follow the evolution of the CHs for the longest possible period of time we produced movies 
from full-disk images obtained with EIT in the 195~\AA\ passband.  An important issue analysing these 
images was the projection effect of loops at the CHBs which can mimic changes which are not real. 
To avoid this problem we have chosen images showing the CHs with the eastern and western 
boundaries not further than -300\arcsec\ and 300\arcsec\ from the disk center, respectively.  
The movies can be seen online as movie\_1996.mp4 (CH1) and movie\_1999.mp4 (CH2). 
The animated sequences show derotated images of the coronal holes with their boundaries 
outlined with a solid line. A dashed-dotted line contour corresponds to the first image of the sequence. 

Following the CHs boundaries evolution during 48 hrs we see a significant dynamics 
related to the evolution of the small-scale loops known as BPs. BPs represent  
$\approx$10\arcsec--40\arcsec\ features with enhanced intensity. Their appearance  is
subject of the spatial resolution of the instrument, so they are often seen as a bright core 
surrounded by a diffuse cloud.  High-resolution observations (1\arcsec-2\arcsec) show 
that BPs consist of several small loops \citep{shgol79, ugarte04, suarez08}. Only BPs 
which are close to the CHBs play a role in the expansion or contraction of the CHs We do 
not identify any flaring events related to this evolution. For the narrow CH1, the emergence, 
evolution and disappearance of BPs results in some cases in a complete 
 displacement of parts of the CH in either  eastern or western direction (see the CH1 movie).  
 \citet{madj04} analysed SUMER  spectral lines taken along a part of the CH1. The authors 
 found numerous transient features called explosive events located along the CH boundaries 
 as well as around the BP which is inside the CH (Fig.\ref{fig1}).
  We also observe displacements of the boundaries which are not related to any well
distinguishable structure in EIT 195~\AA\ or TRACE 171~\AA. A check on temporally close 
images taken in  the EIT He~304~\AA\ (T~$\sim$~4~10$^4$~K) channel shows clearly 
the existence of a BP which apparently has a low temperature and therefore cannot be 
seen in the hotter passbands. The time scale of the changes is defined by the lifetime 
of the BPs which is around 20~{\rm  hrs} in EUV at coronal temperatures \citep{zhang01} 
and 8~{\rm hrs} in X-rays \citep{golub74}. 
 
The next step in our analysis was to study individual cases of evolving loop structures (BPs) 
and their contribution to the coronal hole evolution  in more detail. The analysed FOV 
for CH1 is shown in Fig.~\ref{fig2} presented by two images 
taken at two different times. The two overplotted rectangular boxes (a and b) are shown 
enlarged in Figs.~3 and 4. The two arrows (A and B) on Fig.~\ref{fig3} are pointing at 
loop structures (i.e. BPs) whose evolution led to a change of the boundary. In Fig.~\ref{fig3} 
the arrow A shows the disappearance of a BP associated with
an expansion of the coronal hole while the arrow B is pointing at the appearance 
of a loop structure which led to a contraction of the CH. These changes 
correspond to the size of the evolving BPs. In Fig.~\ref{fig4} another BP evolution 
is followed by a contraction of the CH. 

The CH2 is presented in Fig.~\ref{fig5} by images taken at almost regular time 
intervals (e.g  01:12, 02:00, 02:50, 03:38, 05:15 and 06:49~UT). The two over-plotted 
boxes (a and b) have an enlarged FOV shown in Figs.~6 and 7. 
In Fig.~\ref{fig6} one can see that a formation of a new BP (arrow A) led to the contraction 
of the CH while the disappearance (arrow B) again expanded the CH.  In Fig.~\ref{fig7}
 the disappearance of two BPs is followed by a large increase of the CH region.

\section{Discussion and conclusions}

The major objective of our study was to determine the dynamics of coronal hole boundaries as seen in the
EUV spectral range. To our knowledge similar studies have only been made in the X-ray spectral domain 
\citep{kahler90, kahler02}  which is sensitive to plasmas emitting 
at temperatures above 3~MK. The examples selected for this study represent an equatorial 
extension of a polar  coronal hole  which existed for 4 solar rotations  (during the studied 
rotation the connection was closed above 40\degr) and an `isolated'  coronal hole. Although the 
coronal holes seem to maintain their general shape during a few solar rotations, a closer look 
at their day-by-day and even hour-by-hour evolution demonstrates  a significant dynamics. 
We found that small-scale loops (30\arcsec--40\arcsec) which are abundant 
 along coronal hole boundaries at temperatures up to T $\sim$1.2--1.4 $\times$ 10$^6$~K have 
a major contribution to the evolution of coronal holes. The loops (BPs) emergence, evolution and 
disappearance lead to a continuous expansion or contraction of the coronal holes. In some cases 
that can result in closing  a narrow coronal hole or a large shift of the entire CH.  
These changes appear to be random probably defined by photospheric processes such as convective 
motions, meridional flow,  differential rotation, emergence of magnetic flux as well as intensive inflow of 
unipolar  magnetic flux from attached active region(s).  That most probably triggers intensive magnetic 
reconnection of the closed magnetic field lines of the quiet Sun and the open ones of  the coronal hole. 
The spectroscopic study of CH1 by \citet{madj04} found evidence for magnetic reconnection processes 
happening at transition region temperatures along the CH1 boundaries which strongly supports 
the results of the present work. 

\begin{figure}[htp!]
\centering

        \includegraphics[scale=0.8]{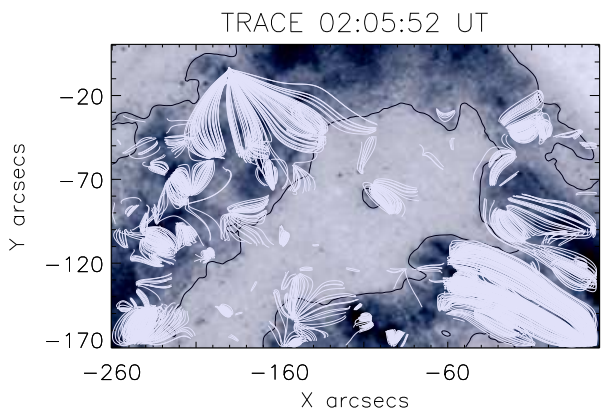}
	\caption{TRACE 171 \AA\ images with a field-of-view as shown in Fig.~\ref{fig5} with 
	over-plotted only closed magnetic field lines connecting magnetic polarities above 20~{\rm G}.  
	The magnetic field lines were obtained with a potential magnetic field extrapolation \citep{thomas2005}.}
		\label{fig8}
\end{figure}

How does the appearance and disappearance of BPs transform the boundaries of the 
coronal holes?  We know so far that 88\% of X-ray BPs are associated with converging 
magnetic polarities \citep{webb93} and most of the BPs are related to cancelling magnetic features.  
Usually one of the magnetic polarities involved is stronger than the other one \citep{madj03} and 
after the full cancellation of the weaker polarity the remnant one either joins the CH dominant polarity 
(if it is of the same sign), expanding the CH surface, or forms a new connection (closed magnetic field lines) 
which triggers contraction of the CH boundaries. This scenario will be tested by applying 
magnetic field extrapolation on Michelson Doppler Imager data in high-resolution mode  for CH2 as well as new 
forthcoming data from SOT/Hinode. An example with potential field extrapolations from SOHO/MDI is
shown in Fig.~\ref{fig8}. In future we are planning to study the
connection between the magnetic and thermal boundary of coronal
holes in more detail with the help of extrapolations from full
disk magnetograms from SDO/HMI, which will provide the global topology
of the coronal magnetic field and allow to distinguish clearly between
globally open and closed regions. The temporal evolution of the fine
structure
at the coronal hole boundary can be studied with higher resolution and
good signal to noise vector magnetograms
from Hinode/SOT. In principal it would be also useful to use more
sophisticated
magnetic field models as the potential field used here. It is well known
that the nonlinear force-free approach is superior for modelling of active
regions,
but unfortunately this approach might not be totally appropriate for coronal
hole
and quiet sun regions, due to the finite beta plasma in the quiet sun
\citep{schrij2005}. Consequently one should model the magnetic field
and plasma self-consistently, in lowest order with a magneto-hydro-static
model
as described in \cite{wieg2006}. Such an approach is also assumed
to
automatically provide consistency between thermal and magnetic boundaries of
coronal holes.  We believe that possible differences between thermal (as
identified
in EUV images) and magnetic hole boundaries might be an artefact of using to
simple magnetic field models.


How do our results contribute to the understanding of the mechanism maintaining the rigid rotation of 
equatorial extension of coronal holes? The rotation of coronal holes is usually determined by measuring 
their centers during several rotation e.g. determine their Carrington longitude from rotation to rotation
(Navarro-Perralta \& Sanchez-Ibarra 1994). These measurements, however, do not take into account the 
small-scale evolution of coronal holes. Our analysis shows that especially in the case of narrow 
coronal holes the open magnetic flux could be totally `recycled' \citep{close2004,close2005} by the continuous interaction with 
the closed magnetic flux and therefore in the following rotation the appearance of the coronal hole will be 
the result of complex changes. An example is the CH1 which almost completely disappeared in 
September 1996 and reappeared again in August 1996.  

New studies  based on specially designed  and  already carried out observing programs with SUMER/SoHO, 
EIS, XRT and SOT/Hinode and  TRACE  during the last two years  are ongoing  which will present the XRT/Hinode 
view  (paper II) and EIS and SUMER view (paper III). These works should reveal more on the physical nature of 
the reported changes  as well as a possible relation between the activity along the coronal hole boundaries and 
the origin of the slow solar wind.  

\begin{acknowledgements} MM thanks ISSI, Bern for the support of the team ``Small-scale transient phenomena 
and their contribution to coronal heating''. Research at Armagh Observatory is grant-aided by the N.~Ireland Department of
Culture, Arts and Leisure. We also thank STFC for support via grants
ST/F001843/1 and PP/E002242/1. The work of TW was supported by DRL-grant 50 OC 0501. 
SoHO is a mission of international collaboration between ESA and NASA.

\end{acknowledgements}

\bibliographystyle{aa}

\end{document}